\documentclass[review]{elsarticle}

\usepackage{lineno,hyperref}
\modulolinenumbers[5]

\usepackage{color}
\usepackage{amsmath}
\usepackage{verbatim}
\usepackage{wasysym}

\journal{Physica E}

\begin{document}

\begin{frontmatter}

\title{Transient dynamics of spin-polarized injection in helical Luttinger liquids}

\author[rvt]{A.~Calzona}
%\ead{cvr@river-valley.com}
\author[focal]{M.~Carrega}
%\ead{kaveh@river-valley.com}
\author[focal]{G.~Dolcetto\corref{cor1}}
\ead{giacomo.dolcetto@gmail.com}
\author[rvt,focal]{M.~Sassetti}
%\ead[url]{http://www.elsevier.com}

\cortext[cor1]{Corresponding author}
%\cortext[cor2]{Principal corresponding author}
%\fntext[fn1]{This is the specimen author footnote.}
%\fntext[fn2]{Another author footnote, but a little more longer.}
%\fntext[fn3]{Yet another author footnote. Indeed, you can have any number of author footnotes.}
\address[rvt]{Dipartimento di Fisica, Universit\`{a} di Genova, Via Dodecaneso 33, 16146 Genova, Italy}
\address[focal]{CNR-SPIN, Via Dodecaneso 33, 16146 Genova, Italy}


\begin{abstract}
We analyze the time evolution of spin-polarized electron wave packets injected into the edge states of a two-dimensional topological insulator. In the presence of electron interactions, the system is described as a helical Luttinger liquid and injected electrons fractionalize. However, because of the presence of metallic detectors, no evidences of fractionalization are encoded in dc measurements, and in this regime the system do not show deviations from its non-interacting behavior. Nevertheless, we show that the helical Luttinger liquid nature emerges in the transient dynamics, where signatures of charge/spin fractionalization can be clearly identified.
\end{abstract}

\begin{keyword}
Topological insulators, Luttinger liquids, transient dynamics
\end{keyword}

\end{frontmatter}

%\linenumbers

\section{Introduction}
The concept of single quasi-particle fails when applied to interacting electrons in one dimension. Indeed, the presence of two distinct Fermi points implies that the low energy excitations are represented by collective charge and spin density waves with bosonic nature \cite{giamarchi03, vondelft98}.
In the presence of electron interactions, a variety of peculiar quantum phenomena emerges. Among them, charge fractionalization represents one of the most striking signature \cite{pam00, deshpande10, steinberg07, kamata14}, being a manifestation of Luttinger liquid (LL) behavior \cite{giamarchi03, tomonaga50, luttinger63}: an electron injected into a LL is fractionalized by interactions into counter-propagating density waves with fractional charges $(1\pm K) e /2$, $e$ being the electron charge and $K$ the Luttinger parameter accounting for the strength of electron interactions ($K=1$ in the non-interacting case). This mechanism is reflected in a sequence of multiple reflections of charge density waves at the {\it interfaces} between interacting and non-interacting regions.

 First evidences of charge fractionalization has been reported in dc tunneling experiments in cleaved-edge-overgrown samples, exploiting momentum-resolved tunneling and multi-terminal geometries \cite{steinberg07, lehur08}.
Here from the knowledge of the current asymmetry and the strength of interactions it was possible to determine the degree of fractionalization.
Also carbon nanotubes attracted a lot of interest in this perspective. Indeed they represent the archetype of non-chiral LLs, characterized by two counter-propagating modes at the Fermi energy. Since their non-Fermi liquid behavior was observed in a variety of transport experiments, in good agreement with theoretical predictions, they have been proposed as ideal playgrounds to explore the phenomenon of charge fractionalization \cite{bena01, trauzettel04}.
Alternative setups can be created by exploting the chiral edge states of integer quantum Hall bars: if the boundaries of the bar are close enough, electron interactions between the counter-propagating edge states are not negligible, and the full system can be viewed as a non-chiral LL \cite{berg09, Horsdal11, Grenier13, Ferraro14}.
Most of the theoretical works studying charge fractionalization have focused on noise measurements in carbon nanotubes and quantum Hall bars. However, from an experimental point of view, clear signatures of fractionalization in noise experiments are still lacking. Indeed, in general, only high-frequency noise carries information about fractionalization, whose detection can be hardly achievable.\\
Very recently, time-resolved experiments have confirmed the physical picture of charge density waves being fractionalized at the interfaces between non-interacting and interacting regions, using integer quantum Hall channels \cite{kamata14, perfetto14}.
To perform such experiments two ingredients are crucial: time-resolved measurement and absence of inelastic scattering. The former allows to follow the dynamic evolution of the charge density waves propagating throughout the system, while the latter guarantees that the phase coherence is preserved.

As far as the first issue is concerned, current measurements with time resolution of $\sim 1$ ps have been performed \cite{kamata14, perfetto14, kumada14}, allowing to detect the real-time dynamics of edge plasmons \cite{ashoori92}. 
On the other hand, to overcome the problem of dephasing, topologically protected edge states can be used, these being characterized by long ($>\mu$m) coherence lengths.

Here we {investigate } evidences of fractionalization and LL physics in a new paradigm of the one-dimensional world: the helical Luttinger liquid (hLL)\cite{wu06, schmidt11, Dolcini12}.
This state can be created at the edge of a two-dimensional topological insulator (2DTI), where a pair of counter-propagating one-dimensional channels appears \cite{qi11, hasan10}. Crucially, the chirality of each channel is intimately connected to its spin-polarization, meaning that in a helical liquid spin-up and spin-down excitations counter-propagate, giving rise to the so-called spin-momentum locking \cite{bernevig06}.
After their discovery in HgTe/CdTe quantum wells (QWs) \cite{konig07}, evidences of helical edge states were found also in InAs/GaSb QWs \cite{liu08, knez11} and other 2D materials \cite{liu11}.
Very recently, the hLL has been observed in a InAs/GaSb QW, where non-Fermi liquid transport properties have been measured \cite{Li15}.
In the presence of time reversal symmetry elastic backscattering from one channel to the counter-propagating one is inhibited \cite{qi11} and non-local transport properties have been experimentally confirmed in multi-terminal geometries \cite{roth09}. Helical edge states offer a promising platform to study quantum phenomena in one dimension, ranging from topological superconductivity \cite{qi08, fu09}, majorana fermions \cite{mi13}, and spin textures \cite{dolcetto13, dolcetto13b}. For example, in hLL, because of spin-momentum locking, one can exploit a spin-polarized tip to inject right-moving or left-moving electrons in the system simply by adjusting the spin-polarization of the tip \cite{das11, dolcetto14}.
Then, because of interactions, the injected electrons fractionalize and propagate into the system in the form of charge and spin density waves.
Remarkably, helicity bounds the charge and spin degrees of freedom together \cite{liutrauzettel11}, so that charge fractionalization is intimately linked to spin fractionalization \cite{das11}. Therefore, if spin-polarized non-interacting electrons enter an hLL, both charge and spin fractional collective excitations are created.

Injection from a spin-polarized tip into an hLL has already been studied \cite{das11}: it was shown that, because of interactions, the injection of spin-up, {\it i.e.} right-moving, electrons induces fractional charge and spin excitations propagating in both the two directions, in sharp contrast with the non-interacting case.
Fractional excitations can be detected via current measurements
which, however, are in general performed via metallic contacts used as detector, whose presence drastically affects the behavior of the whole system \cite{buttiker85, blanter98, buttiker00, gramespacher97, safi95, maslov95, safi99, tarucha95}.

Here we study how the presence of metallic contacts affects the visibility of fractionalization phenomena in transport measurement in hLL. In agreement with previous works on standard LL \cite{lehur08, safi95, maslov95, safi99}, we confirm that dc currents collected at the detectors do not encode information about neither charge nor spin fractionalization, thus apparently preventing the observation of this effect.
A naive explanation of this phenomenon relies on the observation that, because of charge conservation, all the injected charges of a right (left) moving wave packet will be collected at the right (left) contact. Therefore, to observe fractionalization phenomena in hLL physics, alternative detection schemes {have to be} conceived.
Hybrid setup based on topological insulators and capacitive charge sensors, involving quantum dots, single-electron-transistors and high-electron-mobility-transistors, have been proposed, in order to {avoid} the need of Fermi-liquid contacts \cite{garate12}.

Here we reconsider spin-polarized injection in an hLL coupled with metallic
contacts showing that, by means of time-resolved measurement, evidence of fractionalization can be recovered in the transient regime. The time evolution
of such an inhomogeneous system is peculiar since fractionalization arise not only when electrons are injected from the tip into the hLL, but also when
density waves arrive at the interfaces between the hLL and the contacts. Indeed, the charge/spin excitations which reach the interface between an interacting
and a non-interacting region are partially reflected and partially transmitted \cite{safi95, maslov95, safi99, tarucha95}, leading to multiple and subsequent fractionalizations.

By solving the equation of motion for the collective density fields we predict the time evolution of an injected wave packet, finding clear evidences of charge and spin fractionalizations at the interfaces with the contacts. Furthermore, we study the currents detected at the terminals when the tip is biased.
Although dc currents
don't display any signatures of fractionalization, we demonstrate that it is still possible to extract information about hLL physics and fractionalization by studying the transient dynamics.

The paper is organized as follows. In Sec. \ref{model} we provide the theoretical description of the setup with the inhomogeneous hLL model. In Sec. \ref{time} we introduce the equation of motion approach used to evaluate the space-time evolution of the injected wave packets, present the main results and discuss the experimental feasibility. Finally Sec. \ref{conclusions} is devoted to the conclusions.
\section{Model}\label{model}
\paragraph{The helical Luttinger liquid}
The Hamiltonian density of the interacting helical fermions appearing on the edge of a 2DTI is
$
\hat{\mathcal{H}} = \hat{\mathcal{H}}_0 + \hat{\mathcal{H}}^{(int)}
$ where
\begin{equation}
\hat{\mathcal{H}}_0 = -i v_F  \left( \hat{\psi}^{\dagger}_{\uparrow}\partial_x\hat{\psi}_{\uparrow} - \hat{\psi}^{\dagger}_{\downarrow}\partial_x\hat{\psi}_{\downarrow}\right)
\end{equation}
is the free part ($v_F$ is the Fermi velocity) and, introducing the electron density on each channel $ \hat{\rho}_{\sigma}=\colon \hat{\psi}^{\dagger}_{\sigma}\hat{\psi}_{\sigma}\colon$, 
\begin{equation}
\hat{\mathcal{H}}^{(int)} = \frac{1}{2} \sum_{\sigma=\uparrow,\downarrow} \left(g_{4_\parallel} \hat{\rho}_{\sigma}\hat{\rho}_{\sigma} + g_{2_\perp} \hat{\rho}_{\sigma}\hat{\rho}_{-\sigma} \right)
\end{equation}
takes {into} account the electron-electron interaction via the parameters $g_{4_\parallel}$ and $g_{2_\perp}$, under the assumption of short range interaction \cite{giamarchi03}.
The presence of interactions can be treated exactly within the bosonization formalism \cite{giamarchi03, vondelft98}, which consists in rewriting the electron operator {as} $\hat{\psi}_{\sigma}(x)=e^{-i\sqrt{2\pi}\hat{\phi}_{\sigma}(x)}/\sqrt{2\pi a}$, with $a$ a short distance cut-off. Note that, in writing the previous field operator we have omitted the so-called Klein factors \cite{giamarchi03, vondelft98} which are  irrelevant in this context.
The scalar field $\hat{\phi}_{\sigma}$ describes particle-hole excitations and is directly related to the particle density of the relative channel as
$\hat{\rho}_{\uparrow/\downarrow}(x)=\mp\frac{1}{\sqrt{2\pi}}\partial_x\hat{\phi}_{\uparrow/\downarrow}(x)$. Since the electron density is linear in the scalar fields, $\hat{\mathcal{H}}^{(int)}$ can be straightforwardly diagonalized.
By introducing $\hat{\phi}=\frac{1}{\sqrt{2}}\left (\hat{\phi}_{\downarrow}-\hat{\phi}_{\uparrow}\right ), \hat{\theta}=\frac{1}{\sqrt{2}}\left (\hat{\phi}_{\downarrow}+\hat{\phi}_{\uparrow}\right )$
satisfying $[\partial_x\hat{\phi}(x),\hat{\theta}(x')]={-}i\delta(x-x')$, the total Hamiltonian density assumes the standard form of a hLL \cite{wu06}
\begin{equation}\label{eq:Hll}
\hat{\mathcal{H}}=\frac{v}{2}\left [\frac{1}{K}\left (\partial_x\hat{\phi}\right )^2+K\left (\partial_x\hat{\theta}\right )^2\right ]
,\end{equation}
with
\begin{equation}
v=v_F\sqrt{\left (1+\bar{g}_{4\parallel}\right )^2-\bar{g}_{2\perp}^2}
\end{equation}
the velocity of collective excitations and
\begin{equation}
K=\sqrt{\frac{1+\bar{g}_{4\parallel}-\bar{g}_{2\perp}}{1+\bar{g}_{4\parallel}+\bar{g}_{2\perp}}}
\end{equation}
the interaction Luttinger parameter ($K\leq 1$), with $\bar{g}_{4\parallel(2\perp)}=\frac{g_{4\parallel(2\perp)}}{2\pi v_F}$. In the absence of interactions $K=1, v=v_F$ and the system behaves as a Fermi-liquid, with spin-polarized excitations $\rho_\uparrow(x)$ and $\rho_\downarrow(x)$ propagating to the right and to the left respectively.

In the presence of interactions, spin-up and spin-down density waves no longer represent the chiral excitations, which in turn are given by their superpositions
\begin{equation}\label{eq:frac}
\hat{\rho}_{\pm}(x)=\frac{1\pm K}{2}\hat{\rho}_{\uparrow}(x)+\frac{1\mp K}{2}\hat{\rho}_{\downarrow}(x)
,\end{equation}
with $+ (-)$ excitations propagating to the right (left).

\paragraph{Injection and fractionalization}
What happens when electrons are injected into the interacting system crucially depends on the electron interaction through the Luttinger parameter $K$. Consider for example tunneling of spin-up electrons from a nearby tip. In the absence of interactions, spin-up collective excitations are created in the liquid, which, due to spin-momentum locking, propagate to the right with velocity $v_F$. On the other hand, in the presence of electron interactions both right-moving and left-moving collective excitations are created, as shown in Eq. (\ref{eq:frac}), so that a fraction $(1+K)/2$ of the injected flow propagate to the right with velocity $v$, while a smaller amount $(1-K)/2$ propagate to the left with velocity $-v$.
Therefore the presence of electron interactions strongly affect the physical observables (both spin and charge), due to the fractionalization mechanism.

In particular, one could be tempt to conclude that, by measuring the current at the left and the right side of the injection point, clear evidence of charge and spin fractionalizations could be accessed: the measurement of a current at the left of the injection point when the tip is spin-up polarized seems to be a conclusive manifestation of the presence of electron interactions. Unfortunately, this naive expectation is made much more complicate by the presence of metallic contacts in a real measurement.
\paragraph{The role of metallic contacts}
Experimentally, metallic contacts must be attached at some points on the edge, so that the current carried by the interacting helical edge states can be measured. These contacts are macroscopic objects behaving as non-interacting Fermi-liquid.
A standard way to theoretically keep into account their presence is by means of the so-called inhomogeneous Luttinger model \cite{safi95, safi99, dolcini05}.
Here we apply this model to the case of a helical liquid. It corresponds to model the Fermi-liquid contacts as one-dimensional systems with vanishing interactions: formally, one assumes that the interaction parameters $g_{4\parallel}$ and $g_{2\perp}$ are non-vanishing for $|x|<L/2$ only, with $L$ the distance between the contacts, while the interaction is absent for $|x|>L/2$.
\begin{figure}[!ht]
\centering
\includegraphics[width=0.8\textwidth]{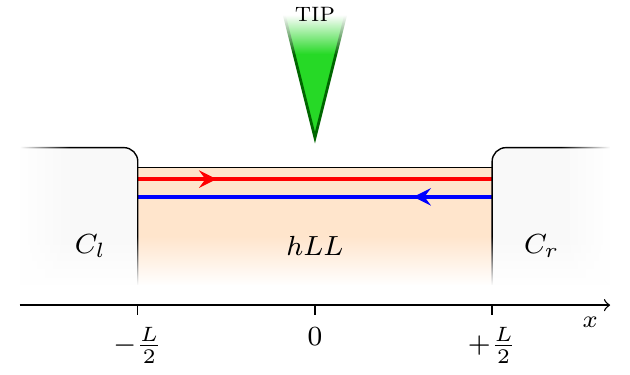}
\caption{(Color online) Schematic view of the setup. A tip injects spin-polarized wave packets inside the hLL ($hLL$), that fractionalize and are detected by the left ($C_l$) and right ($C_r$) contacts.}\label{fig:setup}
\end{figure}
The system, schematically depicted in Fig. (\ref{fig:setup}) (left contact $C_l$, interacting helical liquid $hLL$, right contact $C_r$) is then described by the inhomogeneous hLL Hamiltonian
\begin{equation}\label{eq:Hinh}
\hat{H}=\int~dx~\frac{v(x)}{2}\left [\frac{1}{K(x)}\left (\partial_x\hat{\phi}\right )^2+K(x)\left (\partial_x\hat{\theta}\right )^2\right ]
.\end{equation}
Here the velocity of propagation and the Luttinger parameter acquire a space dependence and are given by:
\begin{equation}\label{eq:parameters}
v(x)=\left \{\begin{array}{cc}
v_F & |x|>L/2 \\ v & |x|<L/2
\end{array}\right ., \ \ \ \ \ \ K(x)=\left \{\begin{array}{cc}
1 & |x|>L/2 \\ K & |x|<L/2
\end{array}\right .
.\end{equation}
In the following we study how the scenario depicted in the presence of spin-polarized tunneling changes due to the presence of the metallic contacts, and {what informations} about hLL physics can be still measured.

\section{Time-resolved dynamics of spin-polarized injection}\label{time}

In the inhomogeneous Luttinger liquid model, the change in the excitations velocities and Luttinger parameters Eq. (\ref{eq:parameters}) makes the interfaces between the hLL and the metallic contacts to effectively behave as potential barriers for the collective excitations: a chiral density wave incoming at one of the interfaces is separated into a transmitted component and into a reflected one.
Note that, because of helicity, both charge and spin density waves undergo scattering phenomena.
Only the transmitted component is finally measured in the contact, while the reflected one propagate toward the other contact, where again it can be either transmitted or reflected back, and so on and so forth.
It is thus important to describe the time evolution of the injected electron wave packet. This can be properly done within the equation of motion approach \cite{perfetto14, maslov95, dolcini05}.

\paragraph{Equation of motion}
The charge density (from now on in units $e$) can be expressed in terms of the scalar field $\hat \phi(x,t)$ as $\hat\rho(x,t)=\hat\rho_\uparrow(x,t)+\hat\rho_\downarrow(x,t)=\frac{1}{\sqrt{\pi}}\partial_x\hat\phi(x,t)$. By means of the continuity equation it is also possible to define the charge current $\hat j(x,t) = - \frac{1}{\sqrt{\pi}}\partial_t\hat\phi(x,t)$.
Note that, because of helicity, charge current and density are related to spin density $\hat{\rho}_s(x,t)=\frac{1}{v(x)K(x)}\hat{j}(x,t)$ and spin current $\hat{j}_s(x,t)=\frac{v(x)}{K(x)}\hat{\rho}(x,t)$ respectively (from now on in units $\hbar/2$). This implies that, in a helical liquid, fractionalization manifests both in the charge and in the spin sector \cite{das11, garate12, calzona15}.

In order to determine the explicit dynamics of the averaged density $\rho(x,t)$ and current $j (x,t)$ it is necessary to solve the equation of motion for the scalar field. Recalling that $\partial_t\hat{\phi}(x,t)=v(x)K(x)\partial_x\hat{\theta}(x,t)$ and $\partial_t\hat{\theta}(x,t)=\frac{v(x)}{K(x)}\partial_x\hat{\phi}(x,t)$, one finds
\begin{equation}\label{eq:motion}
\partial_t^2\hat\phi(x,t)=v(x)K(x)\partial_x\left [\frac{v(x)}{K(x)}\partial_x\hat\phi(x,t)\right ]
,\end{equation}
which can be solved by imposing the continuity of $\hat\phi(x,t)$ and $\frac{v(x)}{K(x)}\partial_x\hat\phi(x,t)$ at the interfaces at $x=\pm L/2$.
In particular, we are interested in studying the dynamic evolution of average density and current after a sudden injection of electrons in the hLL at $t=0$. We thus assign the initial condition, specifying the charge density profile $\rho^{(0)}(x)\equiv \rho(x,t=0)$, with $Q=\int~\rho^{(0)}(x)~dx$ the total injected charge. Electrons are injected from a tip whose spin-polarization forms an angle $\theta$ with the spin quantization axis of the helical fermions \cite{das11}. 
Note that, because of helicity, assigning the initial injected charge density $\rho^{(0)}(x)$ and its spin-polarization (related to the angle $\theta$) is equivalent to assigning initial conditions on both charge and spin densities and currents.
Therefore at time $t=0$ chiral density waves \cite{das11}
\begin{equation}\label{eq:rho0}
\rho^{(0)}_{\pm}(x)=\frac{1\pm K\cos\theta}{2}\rho^{(0)}(x)
\end{equation}
are created in the interacting region\footnote{In the following we consider injections of Gaussian electron wave packets 
$\rho^{(0)}(x) = \frac{Q}{\sqrt{2\pi} \sigma_x} \exp\left[\frac{-x^2}{2 \sigma_x^2} \right]$,
with $Q$ the injected charge and $\sigma_x$ accounting for its spatial distribution, that are localized near the center of the hLL far from the contacts.}.
Note that in the non-interacting case ($K=1$), for $\theta=0$ ($\pi$) only right(left)-moving excitations are created, as expected.

We now consider separately the dynamic of the right-moving chiral component of the injected density and of the left-moving one, see Eq. (\ref{eq:rho0}). Solving the equation of motion Eq. \eqref{eq:motion} with the initial conditions Eq. (\ref{eq:rho0}), we find
\begin{equation}
\label{eq:FrCa:rho2+}
\rho_{{\rm p}}(x,t) =
\begin{cases}
\begin{split}
- \frac{2}{1+K} &\zeta\sum_{n=0}^{+\infty}\gamma^{2n+1} \\ &\rho_{+}^{(0)} \left(-\zeta(x+ v_Ft) + (2n+2)L - \frac{L}{2} \left(\zeta+1\right)\right) \end{split} & x \text{ in } C_l  \\
\begin{split} \sum_{n=0}^{+\infty} \gamma^{2n} &\rho_{+}^{(0)}\left(x- v t + 2n L\right) \\&- \sum_{n=0}^{\infty} \gamma^{2n+1}\,\rho_{+}^{(0)}\left(-x- v t+(2n+1)L\right) \end{split} & x\text{ in } hLL\\ \begin{split}
\frac{2}{1+K}&\zeta \sum_{n=0}^{+\infty} \gamma^{2n} \\ &\rho_{+}^{(0)}\left(\zeta(x- v_Ft) + 2nL-  \frac{L}{2} \left(\zeta-1\right) \right)\end{split} & x\text{ in } C_r.
\end{cases}
,\end{equation}
\begin{equation}
\label{eq:FrCa:rho2-}
\rho_{\rm m}(x,t) =
\begin{cases}
\begin{split}
\frac{2}{1+K}&\zeta \sum_{n=0}^{+\infty} \gamma^{2n} \\ &\rho_{-}^{(0)}\left(\zeta(x+ v_Ft) - 2nL+  \frac{L}{2} \left(\zeta-1\right) \right)\end{split}
 & x \text{ in } C_l  \\
\begin{split} \sum_{n=0}^{+\infty} \gamma^{2n} &\rho_{-}^{(0)}\left(x+ v t - 2n L\right) \\&- \sum_{n=0}^{\infty} \gamma^{2n+1}\,\rho_{-}^{(0)}\left(-x+ v t-(2n+1)L\right) \end{split} & x\text{ in } hLL\\ 
\begin{split}
- \frac{2}{1+K} &\zeta\sum_{n=0}^{+\infty}\gamma^{2n+1} \\ &\rho_{-}^{(0)} \left(-\zeta(x- v_Ft) - (2n+2)L + \frac{L}{2} \left(\zeta+1\right)\right) \end{split} & x\text{ in } C_r.
\end{cases}
,\end{equation}
with $\gamma=\frac{1-K}{1+K}$ and $\zeta = \frac{v}{v_F}$. In the above expression $\rho_{\rm p}(x,t)$ and $\rho_{\rm m}(x,t)$ represent the evolved density profiles associated to the initial chiral density $\rho^{(0)}_+(x)$ and $\rho^{(0)}_-(x)$ respectively. Noteworthy $\rho_{\rm p}(x,t)$ and $\rho_{\rm m}(x,t)$ are no more chiral since they take contributions from all the multiple reflections with the contacts. The time evolution of the total charge density is then given by
\begin{equation}
\label{eq:rho}
\rho(x,t)=\rho_{\rm p}(x,t)+\rho_{\rm m}(x,t).
\end{equation}
Once the excitations have entered the contacts, they propagate with velocity $v_F$. The currents in the left ($j_l$) and in the right ($j_r$) contacts, moving away from the hLL region, can thus be calculated as 
\begin{align}
\label{eq:jl} j_{l}(x,t)&=  v_F  \rho(x,t)\quad & x &\in C_l\\
\label{eq:jr} j_{r}(x,t)&=  v_F \rho(x,t)\quad & x &\in C_r
\end{align}
where, for sake of definition, currents entering in the contacts have positive sign. 

\begin{figure}[!ht]
\centering
\includegraphics[scale=0.3]{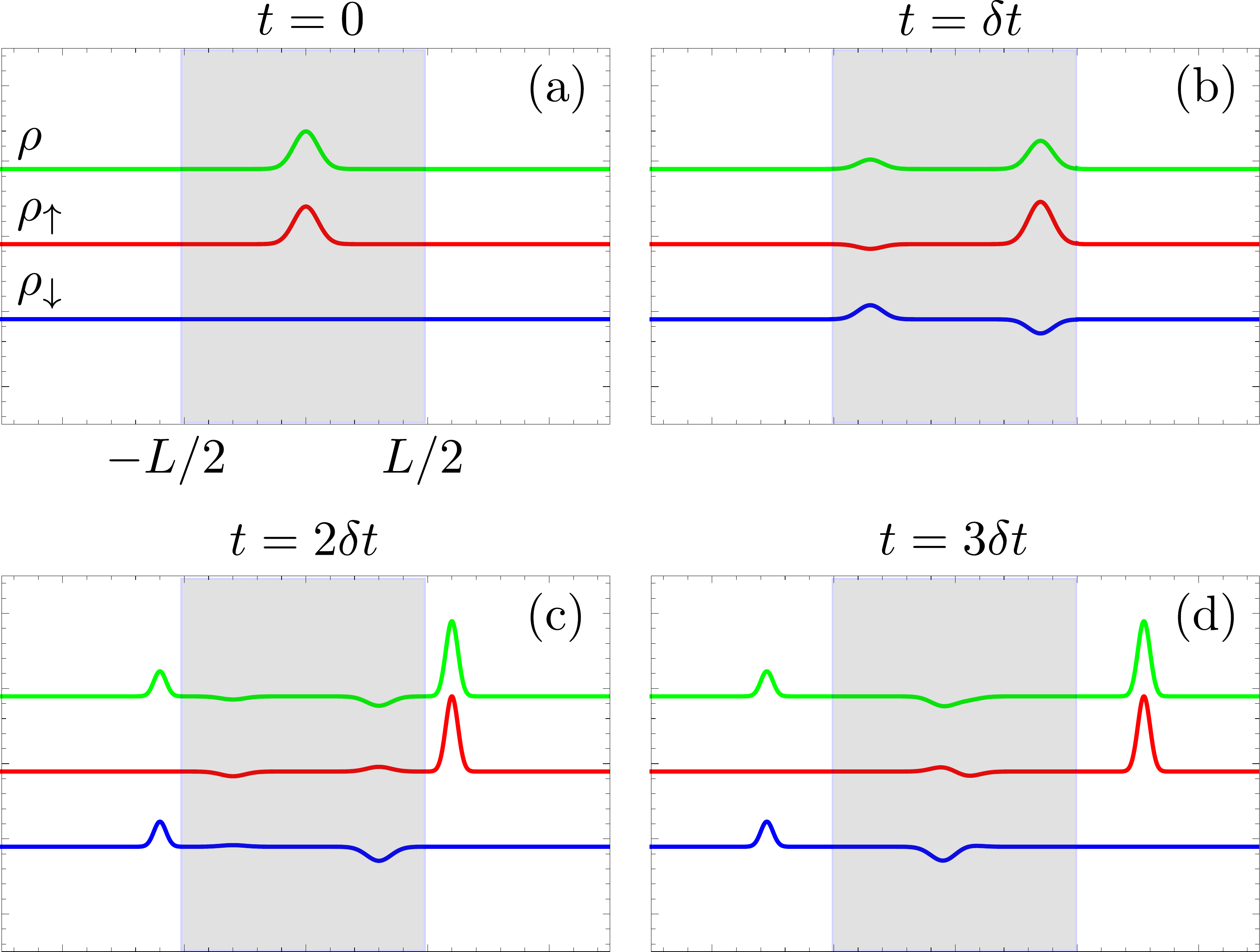}
\caption{(Color online) Charge density $\rho$, spin-up polarized density $\rho_{\uparrow}$ and spin-down polarized density $\rho_{\downarrow}$ profiles at different times. (a) At $t=0$ a spin-up polarized wave packet is injected in the center of the hLL. (b) The fractionalized density waves propagate away from the center of the system toward the contacts. (c) At the contacts they can be either transmitted, and measured, or reflected. (d) The reflected density waves propagate back toward the center of the system. Parameters are: $\theta=0, K=0.5, v_F=10^5$ m/s, $v=v_F/K$, $L=20 \mu$m, $\delta t=35$ ps.}\label{figrho}
\end{figure}

In Fig. \ref{figrho} we report the evolution of a Gaussian electron wave packet injected in the center of the hLL, see Fig. \ref{figrho}(a), from a tip spin-polarized in the $\hat{z}$ direction ($\theta=0$).
Because of interactions, the injected spin-up charge $Q$ splits into two chiral excitations with charge $\frac{1+K}{2}Q$ and $\frac{1-K}{2}Q$, propagating toward right and left contacts respectively, see Fig. \ref{figrho}(b).
At the interfaces, some incident charge is transmitted to the contacts, see Fig. \ref{figrho}(c), where, according with Eqs. (\ref{eq:jl}, \ref{eq:jr}), it can be measured as a current signal.
The reflected density waves propagate toward the opposite contact (Fig. \ref{figrho}(d)), where, again, they are partially transmitted and partially reflected.

\begin{figure}[!ht]
\centering
\includegraphics[scale=0.3]{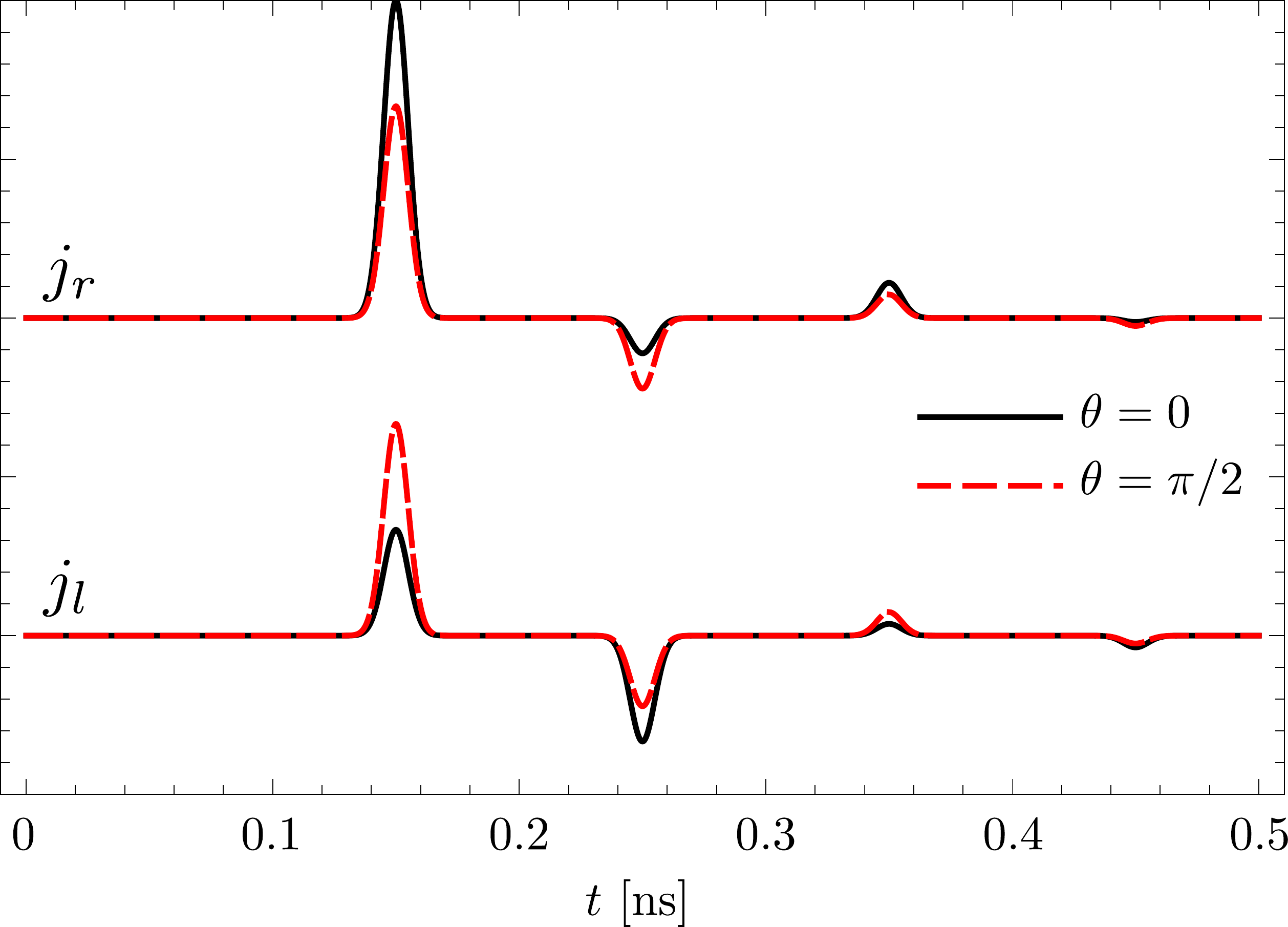}
\caption{(Color online) Currents $j_r(x=L,t)$ and $j_l(x=-L,t)$, see Eqs. (\ref{eq:jl}, \ref{eq:jr}), collected at the right and left contacts respectively as a function of time, after that a Gaussian wave packet has been injected in the center of the interacting region at $t=0$. Currents entering the contacts have positive sign. Two different polarizations of the tip are shown: $\theta=0$ (black full) and $\theta=\pi/2$ (red dashed). Parameters are: $K=0.5, v_F=10^5$ m/s, $v=v_F/K$, $L=20$ $\mu$m.}\label{figj}
\end{figure}

Therefore, the fractionalization mechanism can be probed by means of time-resolved current dynamics, as shown in Fig. \ref{figj}, where the currents arriving at the contacts as a function of time are reported.
Here, the contacts measure the different fractions of the injected wave packet after multiple reflections. Consider for example the current measured at the right contact $j_{r}$ as a function of time (similar arguments hold for the behavior of $j_{l}$).
The first peak (at short times) corresponds to the component of the injected wave packet being directly transmitted through the right interface, while the first dip corresponds to the injected wave packet being reflected at the left interface, then transmitted through the right one and finally detected. Multiple reflections, detected at larger times, give smaller contributions, as argued also from Eqs. (\ref{eq:FrCa:rho2+}, \ref{eq:FrCa:rho2-}).

Crucially, evidences {of fractionalization} are lost in simple dc measurements. Indeed, the total charge detected at the contacts $Q_\alpha=\int_0^{+\infty}j_{\alpha}(x,t)~dt$ can be evaluated from Eqs. (\ref{eq:FrCa:rho2+}, \ref{eq:jl}, \ref{eq:jr}) and reads
\begin{equation}\label{eq:Qdc}
Q_l=\frac{1-\cos\theta}{2}Q, \ \ \ \ \ \ Q_r=\frac{1+\cos\theta}{2}Q
,\end{equation}
independently of the Luttinger parameter $K$.
This result already emerges from Fig. \ref{figj}.
If spin-up electrons are injected ($\theta=0$), all the fractionalized contributions sum up to the total injected charge $Q$ in the right contact, while they cancel out exactly in the left one. On the other hand, if the injected charge density is unpolarized in the $\hat{z}$ direction ($\theta=\pi/2$), the right and left contacts collect the same charge $Q/2$, in agreement with Eq. \eqref{eq:Qdc}.
Note that noise measurements represent an alternative to current detection to investigate the fractionalization phenomenon. However, just like dc current measurements mask the Luttinger liquid behavior, the zero-frequency noise does not carry information about fractionalization in the setup \cite{trauzettel04, Crepieux03, Lebedev04}. Therefore, the noise at finite frequency should be investigated, which however could be challenging to measure, since a wide window spectral range at tens of GHz should be resolved.

So far, we have studied the evolution consequent to the sudden injection of a single localized bunch of electrons at $t=0$, described by an initial density $\rho^{(0)}(x)$. However, the formalism we developed can be used also to study the injection of a generic current in the interacting region, with an arbitrary time-dependence $I^{(inj)}(t)$. Let's consider a series ($n=0,1,2, \dots$) of instantaneous injections of packets, each with associated charge $Q_n$ and subsequent time separation $\Delta t$. Exploiting the linearity of the equation of motion (Eq. \eqref{eq:motion}), the overall density at time $t$ is given by
\begin{equation}
\label{eq:subs_inj}
\rho(x,t) = \sum_{n=0}^{m} Q_n \; \tilde\rho(x,t-n~\Delta t)
\end{equation} 
with $m$ the integer part of $\frac{t}{\Delta t}$. The function $\tilde\rho(x,t)$ is the time evolution, obtained by means of Eqs. (\ref{eq:FrCa:rho2+}-\ref{eq:rho}), of an initial wave packet $\tilde\rho^{(0)}(x)$ with normalized shape ($\int~\tilde{\rho}^{(0)}(x)~dx=1$). Identifying the current $I^{(inj)}(n~\Delta t)=Q_n / \Delta t$ and considering the limit $\Delta t \to 0$ one obtains from Eq. \eqref{eq:subs_inj}
\begin{equation}\label{eq:rhotildeOK}
{\rho}(x,t)=\int_{-\infty}^tI^{(inj)}(\tau)\tilde\rho(x,t-\tau)d\tau.
\end{equation}
Note that for a tip biased with a very sharp pulse, $I^{(inj)}(\tau)=Q^{(inj)}\delta(\tau)$, one recovers the previous result for $\rho(x,t)$ given in Eq. \eqref{eq:rho}.

\begin{figure}[!ht]
\centering
\includegraphics[scale=0.25]{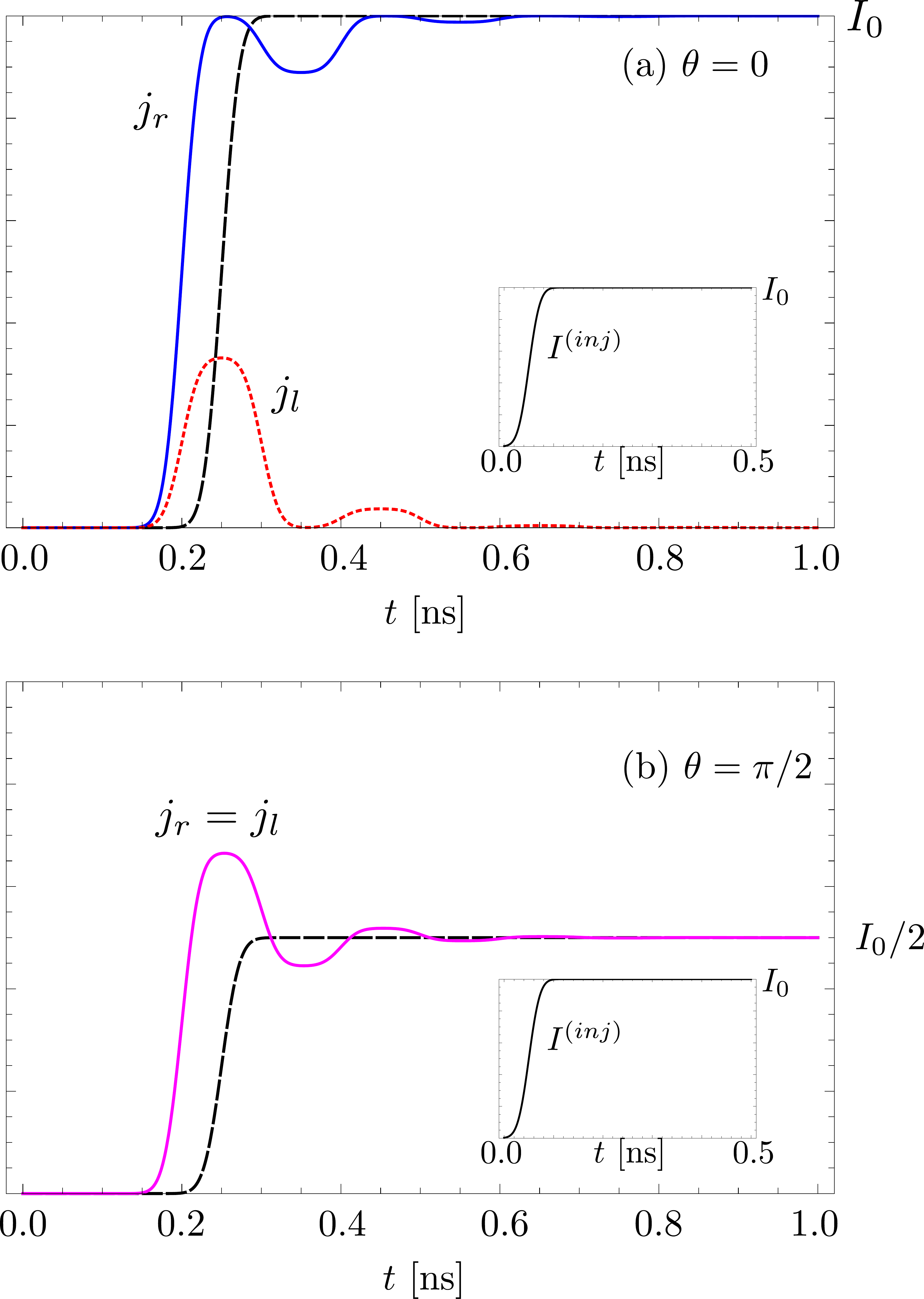}
\caption{(Color online) Currents $j_l$ and $j_r$, see Eqs. (\ref{eq:jl}, \ref{eq:jr}), detected at the left and right contacts respectively due to injection from a biased spin-polarized tip, with the injected current modeled by Eq. \eqref{eq:I(t)} and shown in the insets. Currents entering the contacts have positive sign. We have assumed a narrow tip localized at the center of the hLL ($\tilde \rho^{(0)}(x) = \delta(x)$).
(a) Case $\theta=0$. In the absence of interactions (dashed black), all the injected current flows to the right contact, while no current is detected by the left contact. In the presence of interactions both the right (solid blue) and left (dotted red) contacts detect a non-vanishing current in the transient regime. (b) Case $\theta=\pi/2$. Both the left and right contacts detect an equal current in the absence of interactions (dashed black). However, in the presence of interactions (solid magenta) additional features are observed in the transient regime, due to fractionalization. Parameters are: $K=0.5, v_F=10^5$ m/s, $v=v_F/K$, $L=20 \mu$m, $\Delta \tau=25$ ps}\label{fig3}
\end{figure}

We now use the expression of Eq. (\ref{eq:rhotildeOK}) to study the dynamic evolution of the system in the experimentally relevant case corresponding to injection from a biased tip. We assume that at $t=0$ a constant voltage is imposed to the tip, that starts to inject electron wave packets into the interacting region. Contrary to the previous case of sudden injection, now the tip continuously injects trains of wave packets. The injected current can be modeled as
\begin{equation}
\label{eq:I(t)}
I^{(inj)}(t) = \frac{I_0}{2}\mathrm{Erf}\left (\frac{t}{\Delta \tau}-2\right ),
\end{equation}
with $\mathrm{Erf}$ being the Gaussian errors function and $\Delta \tau$ representing the time interval needed by the tip to be polarized by the battery\footnote{We could have used the simplified model $I^{(inj)}(t)=I_0\Theta(t)$, with $\Theta$ the Heaviside step function, instead of Eq. (\ref{eq:I(t)}), and we would have obtained similar results. However, Eq. (\ref{eq:I(t)}) keeps into account the finite time interval $\sim\Delta \tau$ needed by the tip to be polarized by the battery.}.
The time-dependence of the injected current Eq. (\ref{eq:I(t)}) is shown in the insets of Fig. (\ref{fig3}).
For sake of convenience, we have chosen $\tilde{\rho}^{(0)}(x)$ to be very sharply peaked ($\tilde\rho^{(0)}(x)=\delta (x)$), thus modeling the injection from a narrow tip in the center of the hLL. We underline that the results are not affected if one chooses a broader spatial distribution. 

From Eqs. (\ref{eq:jl}, \ref{eq:jr}, \ref{eq:rhotildeOK}), it is possible to predict the currents measured at the left and right terminals, due to the presence of the injected current $I^{(inj)}(t)$ in Eq. \eqref{eq:I(t)}.
The results are shown in Fig. \ref{fig3}.
Let us analyze the non-interacting case first, represented by the black dashed lines.
Initially ($t\apprle 0.2$ ns), no current is measured by the contacts, since the charge, injected from the tip into the center of the hLL (see the insets in Fig. \ref{fig3}), needs some time $\sim\delta/v_F$ to reach the contacts, with $\delta$ the distance of the detection points from the injection point. In particular, if $\theta=0$, Fig. \ref{fig3}(a), the left contact measures no signal and all the injected current is detected by the right one, while if $\theta=\pi/2$, Fig. \ref{fig3}(b), the injected current is equally detected by both the left and the right contacts ($j_l = j_r$).

{Considering the presence of interactions}  at large times no deviations from the non-interacting behavior are {present}: if $\theta=0$, all the current is detected by the right terminal, while the injected current is equally distributed in the left and right contacts if $\theta=\pi/2$. This result is in agreement with Eq. (\ref{eq:Qdc}): dc measurements cannot provide information about charge fractionalization. On the other hand, differences between the non-interacting and the interacting regimes are captured by \textit{transient} effects.
First, a non zero signal is measured in advance with respect to the non-interacting case, because the time of flight $\sim\delta/v$ is reduced as the velocity is renormalized ($v>v_F$).
The transient regime in the Figure corresponds to $t\apprle 0.6$ ns. Consider the case $\theta=0$ first, corresponding to Fig. \ref{fig3}(a). In this interval, in the presence of interactions the left contact detects a finite current, which vanishes at larger times ($t \apprge 0.6 \, ns$). This result differs from the non-interacting case, where, also in the transient regime, no current is measured by the left contact.
Similar arguments hold for the case $\theta=\pi/2$ shown in Fig. \ref{fig3}(b), where signatures of interactions are found in the transient regime only.
Therefore, sudden injection is not mandatory in order to detect evidences of fractionalization and hLL physics. Indeed, we have shown that these features can be found in the transient regime of ``dc-like'' injection.

These arguments clearly show that, despite dc measurements do not encode information about hLL physics in our setup, time-resolved dynamics and transient effects are able to provide evidences of charge and spin fractionalization.
We remark that our formalism allows to describe a wide range of injection modalities (arbitrary spin-polarization, sudden or adiabatic injection) that allow to investigate fractionalization in hLLs in a variety of setups.

\paragraph{Discussion}\label{discussion}
Finally, we briefly discuss the experimental feasibility of the proposed setup.
The ability to inject localized wave packets into the edge states of a topological system \cite{kumada14,bocquillon14,  waldie15} has been recently improved by Kamata \textit{et al.} \cite{kamata14}, who were also able to perform time-resolved measurements with an accuracy of $\sim 1$ ps.

In order to observe time-resolved dynamics and transient effects, the multiple reflected wave packets must be resolved in time. Let $\sigma_t$ be the  full-width-at-half-maximum of the time distribution of the wave packets (see Fig. \ref{figj}). The time interval between two consecutive wave packets incoming in each terminal is $\sim L/v$, so the condition to resolve two consecutive wave packets is $L\apprge v\sigma_t$. By taking $\sigma_t\sim 20 $ps \cite{waldie15} and $v\sim 10^5 $m/s \cite{konig07}, one finds $L\apprge 2 \mu$m, which represents a lower bound for the possible observation of the fractionalized wave packets. On the other hand, an upper bound is represented by the inelastic mean free path or phase-coherence length $\lambda_{in}$, since inelastic processes can induce backscattering even in the presence of time reversal symmetry, destroying the quantum coherence of the excitations \cite{gusev11, kononov15, olshanetsky15}.
Then, in order to probe the fractionalization mechanism discussed in this work the condition
\begin{equation}\label{eq:exp}
v\Delta t\apprle L\apprle \lambda_{in}
\end{equation}
must be satisfied.
The one-dimensional channels used in Ref. \cite{kamata14} were integer quantum Hall edge states characterized by an inelastic mean free path of the order of $\apprge 100$ $\mu$m. Up to now, in the case of 2DTIs, ballistic transport has been observed in shorter samples only, with an inelastic mean free path of the order of few tens of $\mu$m at best \cite{gusev11, kononov15, olshanetsky15}.
Therefore it may be considered that a $2$ $\mu$m $\apprle L\apprle$ $20$ $\mu$m long hLL should satisfy the requirement Eq. (\ref{eq:exp}).
In order to fulfill Eq. (\ref{eq:exp}) one can work either to reduce the product $v\Delta t$ or to increase the inelastic mean free path $\lambda_{in}$.
On the one hand, InAs/GaSb quantum wells are expected to have edge states with slower propagation velocity ($v_F\sim 10^4$ m/s) with respect to HgTe/CdTe quantum wells \cite{liu08, knez11, garate12}. On the other one, larger value for the inelastic mean free path can be achieved by improving the quality of the samples, thus reducing possible sources of inelastic scattering such as the presence of quasi-2D charge puddles in the bulk \cite{garate12, vaeyrynen13, glazman14}.

\section{Conclusions}\label{conclusions}
In this work we have investigated the fractionalization phenomenon in helical Luttinger liquids,{ where helicity can lead to the fractionalization of both charge and spin degrees of freedom.} 
In particular, we have studied the injection of spin-polarized electrons into the interacting edge state of a 2DTI, contacted with metallic detectors. We have analyzed both the cases of sudden injection, that allows to clearly follow the time dynamics of the injected wave packets, and the case of injection from a constant biased tip. In {both} situations, evidences of non-Fermi-liquid physics are lost in the dc regime, {and} despite the presence of electron interactions, the helical edge states appear as non-interacting at all. However, we have {discussed} how signatures of charge and spin fractionalization can be recovered by studying the time-resolved dynamics of the injection processes. As far as the case of sudden injection is concerned, real-time current detection allows to observe the fractionalized wave packet undergoing multiple reflections. On the other hand, we {have found} evidences of Luttinger liquid physics in the case of injection from a biased tip as well, provided the transient regime is analyzed. We have also {inspected} the experimental feasibility, showing that the proposed setup could in principle be implemented.

\section{Acknowledgments}
We acknowledge the support of the MIUR-FIRB2012 - Project HybridNanoDev (Grant  No.RBFR1236VV), EU FP7/2007-2013 under REA grant agreement no 630925 -- COHEAT, MIUR-FIRB2013 -- Project Coca (Grant No.~RBFR1379UX), the COST Action MP1209 and the CNR-SPIN via Seed Project PGESE003.


\section*{References}

\begin{thebibliography}{77}
\bibitem{giamarchi03} T. Giamarchi, \textit{Quantum Physics in One Dimension} (Oxford University Press, Oxford, 2003).
\bibitem{vondelft98} J. von Delft and H. Schoeller, Ann. Phys. \textbf{7}, 225 (1998).
\bibitem{pam00}K.-V. Pham, M. Gabay, and P. Lederer, Phys. Rev. B {\bf 61}, 16397 (2000).
\bibitem{deshpande10}V. V. Deshpande, M. Bockrath, L. I. Glazman, and A. Yacoby, Nature {\bf 464}, 209 (2010).
\bibitem{steinberg07} H. Steinberg, G. Barak, A. Yacoby, L. N. Pfeiffer, K. W. West, B. I. Halperin, and K. Le Hur, Nature Physics {\bf 4}, 116 (2007).
\bibitem{kamata14}H. Kamata, N. Kumada, M. Hashisaka, K. Muraki, and T. Fujisawa, Nature NanoTech. {\bf 9}, 177 (2014).
\bibitem{tomonaga50} S.-I. Tomonaga, Prog. Thero. Phys. {\bf 5}, 544 (1950).
\bibitem{luttinger63} J. M. Luttinger, J. Math. Phys. {\bf 4}, 1154 (1963).
\bibitem{lehur08}K. Le Hur, B. I. Halperin, and A. Yacoby, Ann. Phys. {\bf 323}, 3037 (2008).
\bibitem{bena01}C. Bena, S. Vishveshwara, L. Balents, and M. P. A. Fisher, J. Stat. Phys. {\bf 103}, 429 (2001).
\bibitem{trauzettel04} B. Trauzettel, I. Safi, F. Dolcini, and H. Grabert, Phys. Rev. Lett. {\bf 92}, 226405 (2004).
\bibitem{berg09}E. Berg, Y. Oreg, E.-A. Kim, and F. Von Oppen, Phys. Rev. Lett. {\bf 102}, 236402 (2009).
\bibitem{Horsdal11}M. Horsdal, M. Rypestol, H. Hansson, and J. M. Leinaas, Phys. Rev. B {\bf 84}, 115313 (2011).
\bibitem{Grenier13}Ch. Grenier, J. Dubois, T. Jullien, P. Roulleau, D. C. Glattli, and P. Degiovanni, Phys. Rev. B {\bf 88}, 085302 (2013).
\bibitem{Ferraro14}D. Ferraro, B. Roussel, C. Cabart, E. Thibierge, G. F\'{e}ve, Ch. Grenier, and P. Degiovanni, Phys. Rev. Lett. {\bf 113}, 166403 (2014).
\bibitem{perfetto14}E. Perfetto, G. Stefanucci, H. Kamata, and T. Fujisawa, Phys. Rev B {\bf 89}, 201413(R) (2014).
\bibitem{kumada14}N. Kumada, P. Roulleau, B. Roche, M. Hashisaka, H. Hibino, I. Petkovic, and D. C. Glattli, ArXiv/Cond. Mat:1407.4379 (2014).
\bibitem{ashoori92}R. Ashoori, H. Stormer, L. N. Pfeiffer, K. Balduin, and K. W. West, Phys. Rev. B {\bf 45}, 3894 (1992).
\bibitem{wu06}C. Wu, B. A. Bernevig, and S.-C. Zhang, Phys. Rev. Lett. {\bf 96}, 106401 (2006).
\bibitem{schmidt11} T. L. Schmidt, Phys. Rev. Lett. \textbf{107}, 096602 (2011).
\bibitem{Dolcini12}F. Dolcini, Phys. Rev. B {\bf 85}, 033306 (2012).
\bibitem{qi11}X.-L. Qi and S.-C. Zhang, Rev. Mod. Phys. {\bf 83}, 1057 (2011).
\bibitem{hasan10}M. Z. Hasan and C. L. Kane, Rev. Mod. Phsy. {\bf 82}, 3045 (2010).
\bibitem{bernevig06}B. A. Bernevig, T. L. Hughes, and S.-C. Zhang, Science {\bf 314}, 1757 (2006).
\bibitem{konig07}M. Koenig, S. Videmann, C. Brune, A. Roth, H. Buhmann, L. W. Molenkamp, X.-L. Qi, and S.-C. Zhang, Science {\bf 318}, 766 (2007).
\bibitem{liu08}C. Liu, T. L. Hughes, X.-L. Qi, K. Wang, and S.-C. Zhang, Phys. Rev. Lett. {\bf 100}, 236601 (2008).
\bibitem{knez11}I. Knez, R.-R. Du, and G. Sullivan, Phys. Rev. Lett. {\bf 107}, 136603 (2011).
\bibitem{liu11}C.-C. Liu, W. Feng, and Y. Yao, Phys. Rev. Lett. {\bf 107}, 076802 (2011).
\bibitem{Li15} T. Li, P. Wang, H. Fu, L. Du, K. A. Schreiber, X. Mu, X. Liu, G. Sullivan, G. A. Cs\'{a}thy, X. Lin, R.-R. Du, Phys. Rev. Lett. {\bf 115}, 136804 (2015).
\bibitem{roth09}A. Roth, C. Brune, H. Buhmann, L. W. Molenkamp, J. Maciejko, X.-L. Qi, and S.-C. Zhang, Science {\bf 325}, 294 (2009).
\bibitem{qi08}X.-L. Qi, T. L. Hughes, and S.-C. Zhang, Nature Physics {\bf 4}, 273 (2008).
\bibitem{fu09}L. Fu and C. L. Kane, Phys. Rev. B {\bf 79}, 161408 (2009).
\bibitem{mi13}S. Mi, D. Pikulin, M. Wimmer, and C. W. J. Beenakker, Phys. Rev. B {\bf 87}, 241405 (2013).
\bibitem{dolcetto13}G. Dolcetto, N. Traverso Ziani, M. Biggio, F. Cavaliere , and M. Sassetti, Phys. Stat. Sol.: RRL {\bf 7}, 1059 (2013).
\bibitem{dolcetto13b}G. Dolcetto, N. Traverso Ziani, M. Biggio, F. Cavaliere, and M. Sassetti, Phys. Rev. B {\bf 87}, 235423 (2013).
\bibitem{das11}S. Das and S. Rao, Phys. Rev. Lett. {\bf 106}, 236403 (2011).
\bibitem{dolcetto14}G. Dolcetto, F. Cavaliere, and M. Sassetti, Phys. Rev. B {\bf 89}, 125419 (2014).
\bibitem{liutrauzettel11}C.-X. Liu, J. C. Budich, P. Recher, and B. Trauzettel, Phys. Rev. B {\bf 83}, 035407 (2011).
\bibitem{buttiker85} M. B{u}ttiker, Y. Imry, R. Landauer, and S. Pinhas, Phys. Rev. B \textbf{31}, 6207 (1985). % Generalized many-channel conductance formula with application to small rings
\bibitem{blanter98}Y. M. Blanter, F. W. J. Hekking, and M. Buttiker, Phys. Rev. Lett. {\bf 81}, 1925 (1998).
\bibitem{buttiker00}M. Buttiker, J. of Low Temp. Phys. {\bf 118}, 519 (2000).
\bibitem{gramespacher97}T. Gramespacher and M. Buttiker,  Phys. Rev. B {\bf 56}, 13026 (1997). 
\bibitem{safi95}I. Safi and H. Schulz, Phys. Rev. B {\bf 52}, 17040(R) (1995).
\bibitem{maslov95}D. L. Maslov and M. Stone, Phys. Rev. B {\bf 52}, 5539(R) (1995).
\bibitem{safi99}I. Safi and H. Schulz, Phys. Rev. B {\bf 59}, 3040 (1999).
\bibitem{tarucha95}S. Tarucha, T. Honda, and T. Saku, Sol. Stat. Comm. {\bf 94}, 413 (1995).
\bibitem{garate12}I. Garate and K. Le Hur, Phys. Rev. B {\bf 85}, 195465 (2012).
\bibitem{dolcini05}F. Dolcini, B. Trauzettel, I. Safi, and H. Grabert, Phys. Rev. B {\bf 71}, 165309 (2005).
\bibitem{calzona15}A. Calzona, M. Carrega, G. Dolcetto, and M. Sassetti, ArXiv/Cond. Mat:1507.06815 (2015), to appear in Phys. Rev. {\bf B}.
\bibitem{Crepieux03}A. Crepieux, R. Guyon, P. Devillard, and T. Martin, {\bf 67}, 205408 (2003).
\bibitem{Lebedev04}A. V. Lebedev, A. Crepieux, and T. Martin, Phys. Rev. B {\bf 71}, 075416 (2004).
\bibitem{bocquillon14}E. Bocquillon, V. Freulon, F.D. Parmentier, J.-M Berroir, B. Plaçais, C. Wahl, J. Rech, T. Jonckheere, T. Martin, C. Grenier, D. Ferraro, P. Degiovanni, and G. Fève, Annalen der Physik {\bf 526}, 1 (2014). 
\bibitem{waldie15}J. Waldie, P. See, V. Kashcheyevs, J. Griffiths, I. Farrer, G. Jons, D. Ritchie, T. Janssen, and M. Kataoka, Phys. Rev. B {\bf 92}, 125305 (2015).
\bibitem{gusev11}G. Gusev, Z. kvon, O. Shegai, N. Mikhailov, S. Dvoretsky, and J. Portal, Phys. Rev. B {\bf 84}, 121302 (2011).
\bibitem{kononov15}A. Kononov, S. Egorov, Z. Kvon, M. Mikhaylov, S. Dvoretsky, and E. Deviatov, JETP Lett. {\bf 101}, 814 (2015).
\bibitem{olshanetsky15} E. Olshanetsky, Z. Kvon, G. Gusev, A. Levin, O. Raichev, N. Mikhailov, and S. Dvoretsky, Phys. Rev. Lett. {\bf 114}, 126802 (2015).
\bibitem{vaeyrynen13}J. I. Vaeyrynen, M. Goldstein, and L. I. Glazman, Phys. Rev. Lett. {\bf 110}, 216402 (2013).
\bibitem{glazman14}J. I. Vaeyrynen, M. Goldstein, Y. Gefen, and L. I. Glazman, Phys. Rev. B {\bf 90}, 115309 (2014).
%********************************
%\bibitem{dagotto13}K. A- Al-Hassanieh, J. Rincon, E. Dagotto, and G. Alvarez, Phys. Rev. B {\bf 88}, 045107 (2013).
\end{thebibliography}
\end{document}